# Mechanical response of all-MoS$_2$ single-layer hetrostructures: A ReaxFF investigation

Bohayra Mortazavi[*,1,≠], Alireza Ostadhossein [2,≠], Timon Rabczuk[1], Adri C. T. van Duin[3]

[1]Institute of Structural Mechanics, Bauhaus-Universität Weimar, Marienstr. 15, D-99423 Weimar, Germany.

[2]Department of Engineering Science and Mechanics, Pennsylvania State Uaniversity, 409A EES Building, University Park, USA.

[3]Department of Mechanical and Nuclear Engineering, Pennsylvania State University, University Park, USA

Molybdenum disulfide (MoS$_2$) is a highly attractive 2D material due to its interesting electronic properties. Recent experimental advances confirm the possibility of further tuning the electronic properties of MoS$_2$ through the fabrication of single-layer heterostructures consisting of semiconducting (2H) and metallic (1T) MoS$_2$ phases. Nonetheless, despite significant technological and scientific interest, there is currently limited information concerning the mechanical properties of these heterostructure systems. This investigation aims to extend our understanding of the mechanical properties of all-MoS$_2$ single-layer structures at room temperature. This goal was achieved by performing extensive classical molecular dynamics simulations using a recently developed RexFF forcefield. We first studied the direction dependent mechanical properties of defect-free 2H and 1T phases. Our modelling results for pristine 2H MoS$_2$ were found to be in good agreement with the experimental tests and first-principles theoretical predictions. We also discuss the mechanical response of 2H/1T single layer heterostructures. Our reactive molecular dynamics results suggest all-MoS$_2$ heterostructures as suitable candidates to provide a strong and flexible material with tuneable electronic properties.

*Corresponding author (Bohayra Mortazavi): bohayra.mortazavi@gmail.com
Tel: +49 157 8037 8770,
Fax: +49 364 358 4511
[≠]Equal contribution

## 1. Introduction

Similar to graphite, multi-layer molybdenum disulfide ($MoS_2$) has been widely used as a lubricant [1] for several centuries. Both graphite and $MoS_2$ crystals are composed of atomic layers with hexagonal lattices that are held together by adhesive interactions through van der Waals forces. However, experimental techniques to fabricate single-layer forms of these structures have been just realized during the recent decades. Graphene [2,3], the single-layer form of graphite presents highest mechanical [4] and thermal conduction [5] properties that out-perform all known materials. Recently, single-layer $MoS_2$ structures have been fabricated successfully with a remarkable prospect for the applications in electrical and optoelectronic devices [6,7] .The main advantage of $MoS_2$ with respect to the graphene lies in its direct-bandgap quasi-two-dimensional semiconducting behaviour, whereas the graphene is categorized as a semi-metal electrically conducting material. Another interesting aspect for $MoS_2$ is its polymorphism nature. A $MoS_2$ sheet is consisting of $S_{top}$–Mo–$S_{Bot.}$ triple atomic planes with strong in-plane bonding [8,9]. Interestingly, depending on the arrangement of S atoms, single-layered $MoS_2$ can exhibit contrasting electronic properties of semiconducting (2H) and metallic (1T) phases [8–11]. Surprisingly, experimental achievements [8] confirm the possibility of fabrication of $MoS_2$ heterostructures made by semiconducting and metallic phases in a single-layer form. These experimental advances along with the theoretical estimations indicate $MoS_2$ as a material with a bright prospect for a wide variety of applications. However, as a material for real applications, a comprehensive understanding of mechanical properties plays a crucial role [12–14]. Mechanical properties of semiconducting $MoS_2$ films have been studied both experimentally [15,16] and theoretically [17–23]. Nevertheless, to the best of our knowledge, mechanical responses of all-$MoS_2$ heterostructures have been studied neither theoretically nor experimentally. This study therefore intends to provide a general understanding concerning the mechanical properties of $MoS_2$ films made from semiconducting and metallic phases.

## 2. Molecular dynamics modelling

The electronic properties of $MoS_2$ strongly correlate with the atomic position of S atoms (Fig. 1a). The 2H phase is the original structure of $MoS_2$ which shows a hexagonal lattice with atomic stacking sequence of ($S_{top}$–Mo–$S_{Bot.}$) ABA (Fig. 1a). On the other hand, 1T phase represents an atomic stacking sequence of ($S_{top}$–Mo–$S_{Bot.}$) ABC, in which the S atoms on the bottom are placed in the hollow center of the hexagonal lattice. To assess the effect of loading direction on the mechanical properties of $MoS_2$ structures, we constructed molecular models along 4 different loading directions. Both 2H and 1T structures present a symmetry angle of 30°. This way, the loading angles of 0 and 30 are commonly called armchair and zigzag directions, respectively. In the present work, we also study the mechanical response of 2H and 1T phases along the loading angles of around 10° (as shown in Fig. 1b) and 20°. In addition, a sample of single layer 2H/1T heterostructure is illustrated in Fig. 1c. In accordance with the experimental observations, in our modeling, the 1T phase inside the 2H phase presents a triangular geometry with a characteristic size, defined by "D". In the present work, we considered different composite structures with various characteristic sizes and concentrations of the 1T phase inside 2H. It should be noted that between 2H and 1T phases, 3 different grain boundaries of "α", "β" and "γ" can be formed as it was experimentally realized by Lin *et al.* [8]. In the present study, we however only considered the "β" grain boundary (shown in Fig. 1c inset). β" grain boundary is also among the common grain boundaries observed in chemically grown 2H polycrystalline $MoS_2$ films [24]. All atomistic models were constructed by observing periodicity along the planar direction to remove the effect of free atoms on the boundaries.

In our molecular dynamics modelling, we used a recently developed ReaxFF [25] bond order potential for introducing the atomic interaction of $MoS_2$ systems. ReaxFF potential is a powerful tool to study a wide range of systems [26–29]. The energy of an atomic system ($E_{system}$) in ReaxFF is described by:

$$E_{system} = E_{bond} + E_{val} + E_{tor} + E_{over} + E_{under} + E_{lp} + E_{vdW} + E_{coulombic} \quad (1)$$

as the sum of the bond energy ($E_{bond}$), valence-angle ($E_{val}$), torsion-angle ($E_{tor}$), lone pair ($E_{lp}$), over-coordinate ($E_{over}$) and under-coordinate ($E_{under}$), energy penalties ($E_{lp}$), plus the nonbonded van der Waals ($E_{vdw}$) and Coulomb ($E_{coulombic}$) contributions. The ReaxFF parameters were trained against a quantum mechanical (QM) data-set that describes bond dissociation and valence angle bending in several small molecules as well as the formation energy and the equations of state of condensed-phases of $MoS_2$ crystalline structures [25]. Mechanical properties were evaluated by performing uniaxial tensile simulation using the *LAMMPS* [30] package.

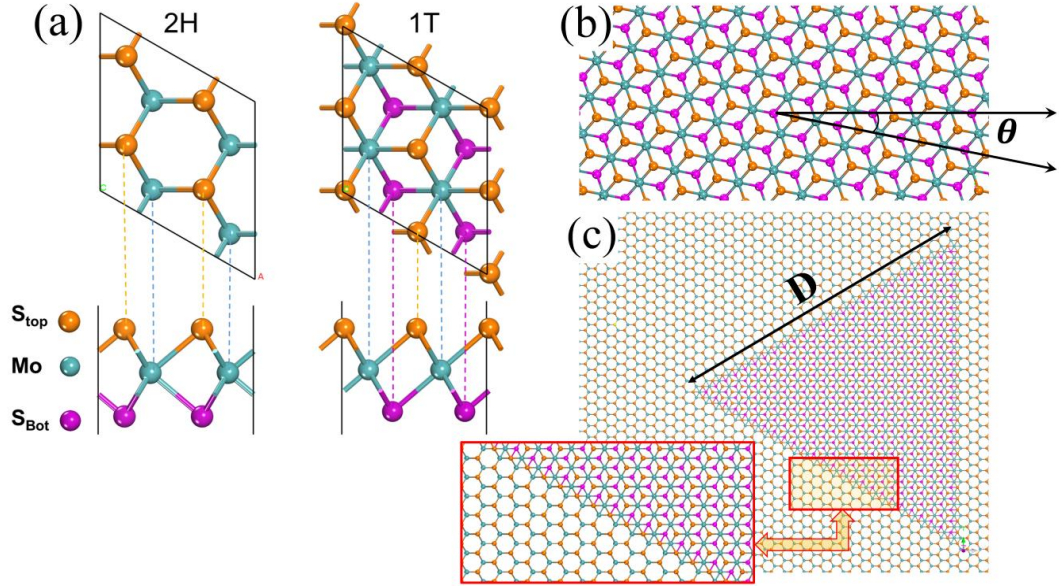

Fig. 1- (a) Schematic illustrations of 2H and 1T single-layered MoS$_2$ phases in basal plane and their corresponding cross-section views. The 2H phase shows a hexagonal lattice with threefold symmetry while in 1T phase the bottom Sulfur (S$_{Bot.}$) is placed in the hollow center of 2H hexagonal lattice. (b) A sample of 1T sheet with a loading angle ($\theta$) of ≈10º with respect to the armchair direction ($\theta$=0.0). (c) A periodic atomistic model of 1T phase inside 2H phase with a triangle domain size (D) of 8 nm.

For the evaluating the mechanical properties of pristine 2H and 1T films, we constructed atomistic models including around 12000 individual atoms in simulation boxes with planar dimensions of about 27 nm × 13 nm. For the 2H/1T heterostructure membranes, we however included more atoms in our modelling, around 20000 atoms in the periodic cells with dimensions of around 24 nm × 24 nm. The time increment of all simulations was fixed at 0.25 fs. We first performed energy minimization before starting the molecular dynamics simulations with a $10^{-6}$ stopping tolerance for energy. Before applying the loading conditions, all systems were relaxed and equilibrated to zero stress at the room temperature using the Nosé-Hoover thermostat (NPT) method with damping parameters of 50 fs and 100 fs for temperature and pressure, respectively. The loading condition was then applied by increasing the periodic box size along the loading direction by a constant engineering strain rate of $1\times10^8$ s$^{-1}$ at every simulation time step. Since atoms on the surface are in contact with vacuum, therefore the stress along the thickness is zero. Thus, in order to ensure uniaxial stress conditions, periodic simulation box along the sample width was altered to reach zero stress in this direction using the NPT method. The strain at each simulation time step was acquired by multiplying the applied engineering strain rate to the time of the loading step. We finally calculated the Virial stresses at every strain level. The stress and strain values were averaged during every 250 fs intervals to report stress-strain curves. The nominal thickness of the single-layer $MoS_2$ was assumed to be 6.1 Å which was obtained based on the concept of bending rigidity [25].

3. Results and discussions

In Fig. 2, the acquired stress-responses for defect-free 2H and 1T phases along four different loading directions at the room temperature are illustrated. In Table 1, we compare our results with experimental and theoretical predictions for $MoS_2$. In this case, we only compare our results for 2H phase because we could not find any information regarding the 1T phase mechanical properties. Interestingly, the predictions by the ReaxFF potential for the elastic modulus and tensile strength are in a remarkable agreement with experimental results. In particular, the calculated elastic modulus falls in an excellent agreement with that of the bulk $MoS_2$ [18] and it is within the lower bond of the experimental tests for few-layer $MoS_2$ films [15,16]. On the other hand, our ReaxFF molecular dynamics modelling propose the 1T phase as a strong material which can yield a mechanical strength as high as almost half of the pristine 2H phase. The elastic modulus, tensile strength and corresponding failure strain of 1T phase were simulated to be 91±7 GPa, 11±1 GPa and 0.22±0.03, respectively.

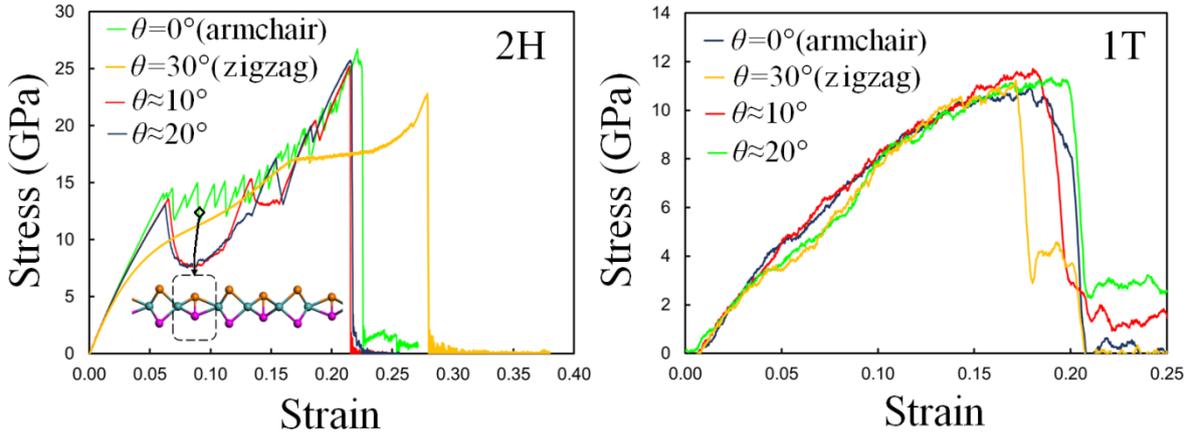

Fig. 2- Calculated uniaxial stress-strain responses of defect-free 2H and 1T $MoS_2$ phases along different loading angles (θ) at the room temperature.

Our results reveal that elastic response of $MoS_2$ phases are dependent on the loading direction and along the armchair $MoS_2$ present slightly higher rigidity than along the zigzag. In addition, in the case of 2H phase our ReaxFF based calculation reveal an unusual pattern in the stress values after passing the first yield point (end of the linear part in stress-strain response). In this case, we found that further increasing of the strain levels from the yield point cause the structure to start to undergo a phase transformation. This phase transformation could be explained by the fact that during the loading condition and stretching of the sheet in the planar direction the atoms along the thickness contract and approach each other. This way the distance between S atoms on the bottom and the top decreases and after a while they tend to form a bond. The formation of S-S bonds is based on the bond-distance cut-off and ReaxFF bond-order of S atoms and it can be considered as a phase transformation (as shown in Fig. 2a, inset). This bond formation helps the specimen to flow easier along the planar direction. Nevertheless, this phase transformation due to the S-S bond formations, occur and extend gradually leading to irregular stress-strain curves.

Table 1, Comparison of present study calculated mechanical properties for $MoS_2$ with previous experimental and theoretical studies.

| Reference | Method | Structure | Elastic modulus (GPa) | Tensile strength (GPa) |
|---|---|---|---|---|
| Present | ReaxFF based molecular dynamics at 300 K | Single-layer 2H | 245±15 | 24.5±2.5 |
| Bertolazzi et al. [15] | Atomic force microscope nano-indentation experiment | Few-layer 2H | 270±100 | 22±4 |
| Castellanos-Gomez et al. [16] | Atomic force microscope nano-indentation experiment | Few-layer 2H | 300±70 | --- |
| Lorenz et al. [17] | First-principles molecular dynamics | Single-layer 2H | 262 | 21 |
| Feldman [18] | Lattice dynamics calculation of dispersion curves | Bulk $MoS_2$ | 240 | --- |
| Zhao et al. [22] | Stillinger-Weber based molecular dynamics at 300 K | Single-layer 2H | 120 | --- |
| Dang et al. [19] | REBO based molecular dynamics at 10 K | Single-layer 2H | 232 | 20.6 |
| Li et al. [23] | First-principles density functional theory calculations | Single-layer 2H | 221±2 | 22±5.5 |
| Ataca et al. [20] | First-principles density functional theory calculations | Single-layer 2H | 240 | --- |

In order to assess the reversibility of the deformation process of pristine 2H $MoS_2$, we performed uniaxial unloading simulations. In this case, several uniaxially loaded films under different strain levels were selected. For each selected sample, we then decreased the simulation box size along the loading direction using the constant strain rate of $10^7 s^{-1}$ and at the same time the stress on the perpendicular direction was controlled by the NPT method to observe zero stress condition in this direction (uniaxial stress condition). In Fig. 3, the loading and unloading stress-strain responses of a single-layer 2H $MoS_2$ along the armchair direction are illustrated. Interestingly, for the strained samples up to the strain of 0.06, the unloading stress-strain curves coincide well with that of the loading curve and the deformation is completely reversible. Such an observation implies that the structure is stretched within the elastic regime and no plastic deformation exist. In this case, the 0.06 strain is the yield point in which the initial linear relation in stress-strain curve ends. As discussed earlier, in this point the S-S bond formations start to occur, resulting in fluctuations in the stress-strain response. As it is illustrated in Fig. 3, for the loaded structures with strain levels higher than the 0.06, the unloading stress-strain curves show an almost linear drop pattern and such that the strain at the zero stress is positive. This means that the deformation is not completely reversible and structure has undergone persistent lattice changes. We found that the S-S bond formations are the main factor resulting in such a behaviour because of the fact that during the unloading and the structural relaxation, this bonds cannot be completely removed.

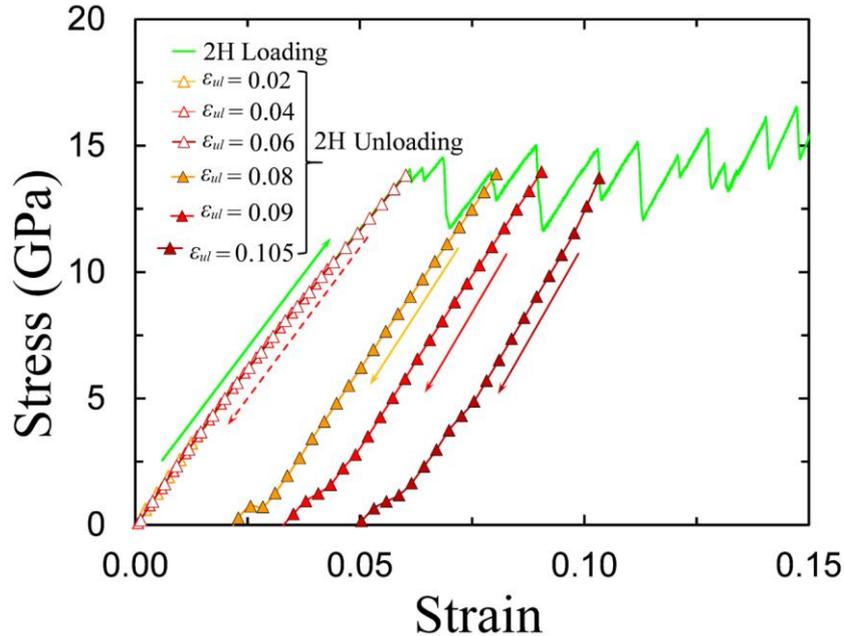

Fig. 3- Loading and unloading stress-strain responses of a single-layer 2H $MoS_2$ along the armchair direction. We selected several strained films under different strain levels ($\varepsilon_{ul}$) and then applied the unloading condition.

Based on the previously reported first principles results, 2H $MoS_2$ phase is more stable than the 1T phase [31,32]. We also used ReaxFF to investigate the structural stability of single-layer $MoS_2$ pristine phases through considering the minimum energy pathways for the 1T↔2H transformation. This transformation was probed by moving one plane of S atoms with respect to the rest of the crystal using the nudged-elastic band (NEB) method, as it is implemented in the *LAMMPS* package. The NEB result for 1T↔2H transformation based on the ReaxFF is illustrated in Fig. 4. According to our calculation, the 2H-to-1T transition energy barrier is obtained to be around 1.1 eV. Based on the ReaxFF, 2H $MoS_2$ is more stable than 1T phase by almost 0.55 eV, which is in a good agreement with the reported value of around 0.6 eV by Guo *et al.* [33]. Fig. 4 also indicates that 1T to 2H $MoS_2$ transition is exothermic and this happens provided that the enough activation energy is supplied, either through some external energy impact or lattice distortion.

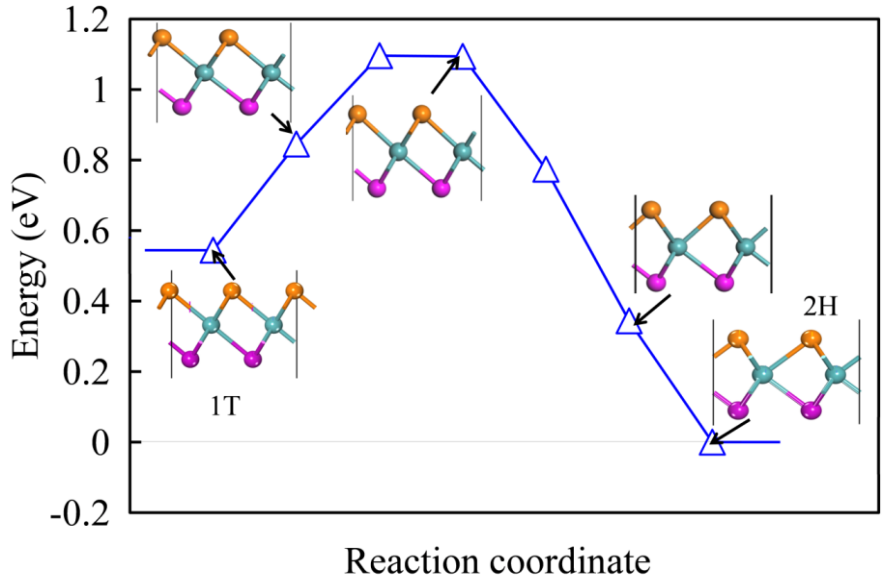

Fig. 4- Nudged-elastic band (NEB) result for the minimum energy pathway for the 1T↔2H transformation.

Like all known materials, experimentally fabricated $MoS_2$ membranes are not expected to be ideal and different types of defects and impurities may exist in their atomic lattices. For example, crystal growth occurring during the chemical vapour deposition (CVD) fabrication of $MoS_2$ sheets lead to the formation of grain boundaries with various types of defects [24,34,35]. The $MoS_2$ films may also include various type of atomic impurities such as oxygen [36–38]. These effects can naturally affect the mechanical properties of $MoS_2$ films. For the theoretical investigation of defects and impurities effects on the mechanical response of MoS$_2$ membranes, ReaxFF reactive molecular dynamics modelling can be considered as a promising approach.

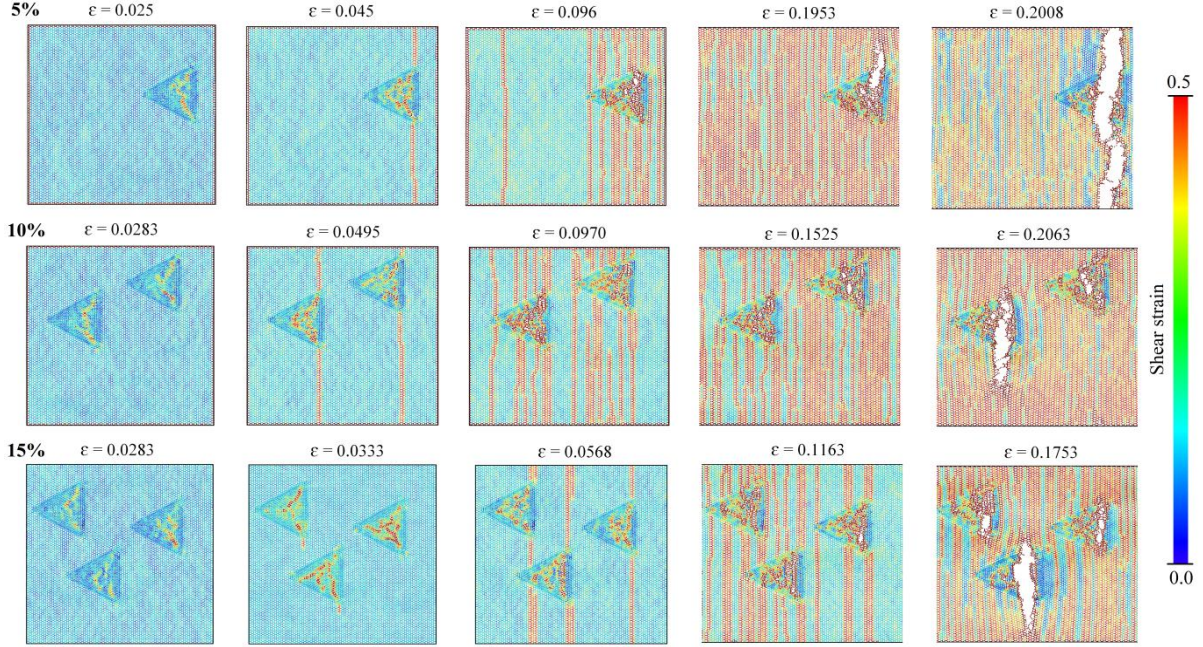

Fig. 5- Deformation processes of single-layer MoS$_2$ heterostructures for different concentrations of 5%, 10% and 15% of 1T phase inside the pristine 2H phase at various stages of the uniaxial tensile strains ($\varepsilon$) at the room temperature. The highlighted atoms with red colors correspond to those with higher strains. OVITO package [39] is used to visualize the *LAMMPS* atomic trajectory outputs.

Next, we study the mechanical properties of 2H/1T single-layer heterostructures. It is predictable that the mechanical response of 2H/1T heterostructures would be lower than the pristine 2H phase. This conclusion is on the basis of the fact that our estimations for the elastic modulus and tensile strength of the 1T phase are almost half of the those for 2H phase. In Fig. 5, we discuss the deformation process of composite structures made by 1T phase with a domain size of 8 nm and with different concentrations. In all studied cases, we could observe that the deformation process is independent of the 1T phase concentration. During the loading condition, we found that at the early stages of the loading the atoms inside the 1T phase were strained more severely, causing structural distortions. These lattice distortions inside the 1T domains were initiated due to the stress concentrations at the sharp corner of the triangle which rapidly extend toward the center of the grains. At higher strain levels, stress concentrations at triangle corners not only induce higher deformations inside the 1T phase but it also stimulate the phase transformation initiation in the 2H phase. The phase transformation later extends gradually throughout the entire 2H phase in the form of parallel lines perpendicular to the loading direction. Nevertheless, the ultimate tensile strength point occurs in a point in which crack forms inside the soft 1T phase. This crack rapidly grow perpendicular of the loading direction and then passes through the triangle corner toward the surrounding 2H phase and finally leads to the sample rupture. Despite the stress concentration at the triangle corners, the crack initiation occurs inside the 1T phase which implies that the 1T/2H grain boundary is remarkably stronger than the original 1T phase.

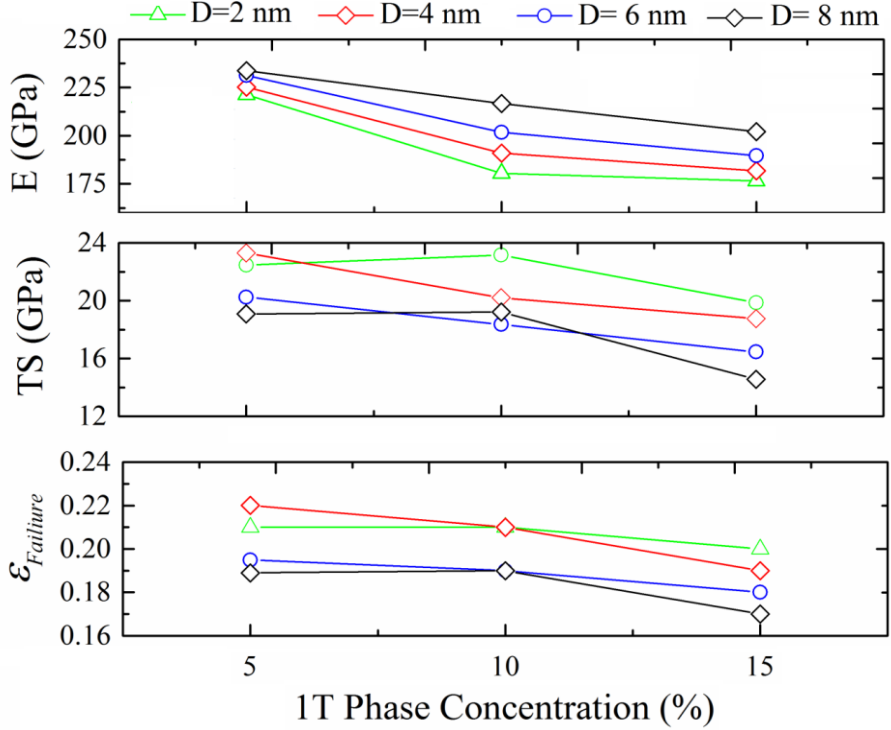

Fig. 6- Elastic modulus (E), tensile strength (TS) and strain at failure ($\varepsilon_{Failure}$) of $MoS_2$ heterostructures as a function of concentration and domain size (D) of 1T phase inside 2H $MoS_2$, at room temperature.

Our results for the mechanical properties of 2H/1T heterostructures are summarized in Fig. 6. As expected, in all cases by increasing the 1T phase concentration the effective elastic modulus, tensile strength and strain at failure decrease. In addition, the mechanical properties of composite structures are predicted to be higher than those for original 1T phase. This finding is due to the presence of strong grain boundaries between 1T and 2H phase. We next study the effects of 1T phase domain size. From a theoretical point of view, by decreasing the domain size, the ratio of atoms along the 2H/1T grain boundaries with respect to the total number of atoms increases. However, since we already found that mechanical strength of atoms along grain boundary is higher than those in defect-free 1T phase, therefore one may conclude that by decreasing the domain size the tensile strength should improve. Nonetheless, by decreasing the domain size the stress concentration also increases. This way, the interplay between these two contrasting factors and the statistical nature of the problem due to the temperature effect define the failure strength and corresponding strain at failure of considered systems. This can relatively explain our molecular dynamics results which suggest that the 1T phase domain size do not present a clear effect on the tensile strength and failure strain of 2H/1T $MoS_2$ heterostructures. On the other hand, our results reveal that the elastic modulus decreases by decreasing the domain size of the 1T phase. This basically means that the adhesion rigidity of the grain boundaries is lower than that for the pristine 2H phase.

## 4. Conclusion

We performed extensive classical molecular dynamics simulations to provide a general viewpoint concerning the mechanical properties of all-$MoS_2$ single-layer heterostructures. To this aim realistic atomistic models were constructed with different concentrations and domain sizes for the metallic phase inside the semiconducting $MoS_2$. Our predictions based on a recently parameterized ReaxFF potential for the mechanical properties of pristine semiconducting $MoS_2$ phase was found to fall within experimental and first principles theoretical values available in the literature. Based on our classical molecular dynamics simulations, the defect-free metallic $MoS_2$ is predicted to yield a mechanical strength and elastic modulus as high as almost half of the pristine semiconducting phase. In addition, we found that by increasing the concentration of metallic phase inside the semiconducting $MoS_2$ the effective elastic modulus, tensile strength and strain at failure of heterostructures decrease. Nevertheless, it is predicted that the mechanical properties of $MoS_2$ heterostructures are yet higher than the properties of original metallic $MoS_2$. It is concluded that the tensile strength of grain boundaries between the semiconducting and metallic $MoS_2$ are stronger than the strength of metallic phase. We however emphasise that our results are based on assuming ideal interface between the metallic and semiconducting phases. During the tensile deformation of $MoS_2$ heterostructures, we found that at the early stages of the loading the atoms inside the metallic phase were strained more severely, which was initiated due to the stress concentrations at the sharp corners. At higher strain levels, stress concentrations at corners not only induce higher deformations inside the metallic phase but it also stimulates the phase transformation in the pristine semiconducting phase. At the ultimate tensile strength point a crack forms inside the soft metallic phase which rapidly grow and leads to the sample rupture. Our modelling results confirm that all-$MoS_2$ heterostructures are remarkably strong and flexible materials.

Acknowledgment

BM and TR greatly acknowledge the financial support by European Research Council for COMBAT project (Grant number 615132).